\documentclass[preprint]{IEEEtran}

\ifCLASSINFOpdf
  \usepackage[pdftex]{graphicx}
  \usepackage{booktabs}
\else
\fi

\usepackage{array}
\usepackage{wrapfig}
\usepackage{graphicx}

\usepackage{url}
\usepackage{hyperref}

\begin{document}

\title{Interconnection between darknets}

\author{\IEEEauthorblockN{Carlos Cilleruelo\IEEEauthorrefmark{1},Luis de-Marcos\IEEEauthorrefmark{2},
Javier Junquera-Sánchez\IEEEauthorrefmark{3}, and
José-Javier Martínez-Herráiz\IEEEauthorrefmark{4}}
\IEEEauthorblockA{ Departamento de Ciencias de la Computación,
Universidad de Alcalá\\
Edificio Politécnico. Campus Universitario, Ctra. Madrid-Barcelona km, 33, 600
Alcalá de Henares. Madrid. Spain\\
Email: \IEEEauthorrefmark{1}carlos.cilleruelo@uah.es,
\IEEEauthorrefmark{2}luis.demarcos@uah.es,
\IEEEauthorrefmark{3}javier.junquera@uah.es,
\IEEEauthorrefmark{4} josej.martinez@uah.es}}

\markboth{Accepted version. Cilleruelo,  \MakeLowercase{\textit{et al.}}: EEE Internet Computing. 1089-7801 ©2020 IEEE. doi.org/10.1109/MIC.2020.3037723}{}

\maketitle

\begin{abstract}
Tor and i2p networks are two of the most popular darknets. Both darknets have become an area of illegal activities highlighting the necessity to study and analyze them to identify and report illegal content to Law Enforcement Agencies (LEAs). This paper analyzes the connections between the Tor network and the i2p network. We created the first dataset that combines information from Tor and i2p networks. The dataset contains more than 49k darknet services. The process of building and analyzing the dataset shows that it is not possible to explore one of the networks without considering the other. Both networks work as an ecosystem and there are clear paths between them. Using graph analysis, we also identified the most relevant domains, the prominent types of services in each network, and their relations. Findings are relevant to LEAs and researchers aiming to crawl and investigate i2p and Tor networks.
\end{abstract}


%
\IEEEpeerreviewmaketitle

\section{Introduction}
Over the last years, the exploration of the darknets has become of great importance for governments and Law Enforcement Agencies (LEAs)~~\cite{chertoff_impact_2015}. Even though the term darknet may refer to different things, it represents the space of the Internet that has been hidden by design ~\cite{owen_Tor_2015}, through encryption or different routing overlapped technologies.

"The Onion Router"\footnote{https://www.Torproject.org/} (Tor) is the most popular technology of the darknet. Tor has been audited and investigated numerous times~\cite{snader2008tune}~\cite{mccoy2008shining}~\cite{loesing2010case}. Tor has over 2000000 daily users and more than 6000  relays~\footnote{Tor Metrics: https://metrics.torproject.org/}. Relays ensure that the network works, and they provide the privacy of the Tor network. The number of existing relays suggests significant support by an active community around this darknet. Another important feature of Tor is that it is possible to create websites and services called hidden services. Hidden services can only be accessed if we are connected to the Tor network. The original principle of Tor's hidden services was to avert censorship and facilitate freedom of speech. However, these hidden services have also turn into a space for criminal activity, like drug trafficking~\cite{dolliver2015evaluating}. Tor popularity increased even more since it can also be used to offer anonymity in operating systems, like Tails. Tails became famous since Edward Snowden recommended using it~\footnote{https://twitter.com/snowden/status/941018955405242369}. 

The second larger darknet is the "Invisible Internet Project"\footnote{https://geti2p.net/}, i2p. Like Tor hidden services, i2p offers eepsites, which are the websites and services available in this darknet. There are no public stats about daily users and services, but i2p is frequently maintained and developed. Also, like Tor hidden services, eepsitees became a place for illegal activities~\cite{wilson2016forensic}~\cite{bazli2017dark}.

To fight illegal activities, LEAs need techniques to discover, investigate, and correlate data between darknets. This paper reports the research and tools for exploring Tor and i2p. We started developing tools for Tor like crawlers and scanners of hidden services. Initial exploration returned numerous references to eepsites. These references lead us to a second phase where we developed new tools to crawl and discover i2p, which also returned references to Tor hidden services sites. Crawling both darknets in parallel also provides more information, because crawlers get feedback from the other darknet. We found 8148 references to i2p eepsites from Tor hidden services and 487 Tor hidden services domains inside i2p eepsites. The information gathered by these tools was used to create a dataset of domains from both darknets and their connections, which provides a map of the darknet. Using graph analysis techniques, we further analyzed the connections between both darknets, and also identified the most relevant actors and types of services offered by each network.

The main objective of this paper is to analyze the interconnection between i2p and Tor, demonstrating the necessity to crawl both networks to get and to study the structure of the darknet and its services. Since the development of Tor and i2p networks is beyond the scope of this paper, the following sections describe the creation of the dataset, the methodology to analyze the graph of the network, the results, and the implications for LEAs. Contributions of this paper are summarized as follows:

\begin{itemize}
    \item A dataset combining i2p and Tor crawled domains and their connections. Connections include links to any domain from both darknets. To our best knowledge, this is the first public dataset reporting i2p domains and also reporting domains from both networks. The dataset is publicly accessible in GitLab, \url{https://gitlab.com/ciberseg-uah/interconection-between-darknets-dataset}
    \item A mapping of the darknet that covers a significant part of it. 
    \item A rank of the most relevant domains in the darknet in terms of their position and influence, which shows that each darknet plays a different role offering specific services, and emphasizes the necessity to crawl and study one darknet to find sites in the other.
\end{itemize}


\section{Background}
There are several studies that report the crawling~\cite{nunes2016darknet} ~\cite{owen_Tor_2015}, discovery~\cite{chaabane2010digging} and dataset creation~\cite{ALNABKI2019212} for the Tor network. Also, several approaches focus on hybrid crawling~\cite{iliou2016hybrid}, searching for data in Tor, i2p, and Freenet. Most of the existing research tries to identify threat intelligence information and other critical information. The approach followed in this paper is based on graph analysis to map the darknets. Our purpose is to evidence the connections between darknets and identify the relevant sizes and services that relate both darknets. To do that, we represent Tor and i2p as a directed graph and we apply graph analysis to measure the interdependence between networks and the position of individual domains. 

Graph theory is useful to analyze social networks~\cite{goodreau2007advances}. Furthermore, existing studies use graph analysis to investigate the Tor darknet~\cite{sanchez2017onions}~\cite{ALNABKI2019212}. This study uses graph analysis to identify relevant hidden services, eepsites, and to study the connection between the Tor and i2p. 

In our work, the detection of influential nodes is done by analyzing the connections between them. This analysis was performed merging the data from i2p and Tor hidden services in a single dataset.

\section{The dasaset of i2p and Tor domains}
In order to study the connections between Tor and i2p, we started by building a dataset of domains and relations. It stores which domain links to others. This returns a dataset that represents a network to study, through graph analysis, the interconnection between darknets. 

At the moment of writing this paper, there is not a public dataset with data from i2p and Tor network. We present here the first dataset with domains from i2p and Tor network. The dataset contains 49249 domains (2687 domains from i2p and 46562 domains from Tor) and 304673 relationships between domains. There are datasets and collections of darknet domains like Ahmia dataset\footnote{https://ahmia.fi/} or DUTA-10k~\cite{ALNABKI2019212}. Lists of domains can also be found in popular websites like Pastebin and Reddit. However, none of them contains relationships between i2p and Tor domains. A comparison of the size of our dataset with existing ones returns that DUTA-10K has 10k onion domains compared \textgreater 46k of our dataset. These figures do not include the fact that our dataset contains i2p domains too. 

However, it is not possible to compare data of i2p since, at the moment of writing this, there are no public datasets. We can compare and check our data with Pastebin and Reddit domain lists, and our list of i2p domains turns out to be larger. Furthermore, an estimation of the number of i2p eepsites is difficult to find in academic literature.

\subsection{Dataset creation}
To create the dataset and obtain all the possible information, we combined different approaches: using open-source lists of domains, crawling darknet sites, generating and verifying new domains, and deploying a modified relay in Tor.  

Firstly, we collected lists of hidden services and eepsites from open sources. There are many domains indexed in open sources like Pastebin or Reddit. Initially, we got the domains from these open sources manually, and then we developed automatic tools to perform this task. 

An additional approach to obtain new domains is crawling the darknets. The specific details concerning how the crawler work is out of the scope of this paper, but we provide a short description. We built a crawler similar to the Ahmia project. Our crawling process started with the Hidden Wiki\footnote{http://zqktlwi4fecvo6ri.onion}, and explored recursively the links found there. After that, we improved the crawler for exploring and storing i2p eepsites. We used a similar approach to crawl i2p. Initially, we browsed i2p eepsites that list services. Using this crawling process, we obtained most of the data of our i2p/Tor dataset. 

It is possible to take advantage of how Tor works to implement another two approaches: domain generation and modification of a Tor relay. Tor provides the possibility of domain generation using known words. A few hidden services need to be easily identified, so they generate domain names using keywords. For example, hidden services focused on selling drugs may try to generate domains with the words \textit{drug} or \textit{market} inside their domain name. In order to find more Tor domains, we developed a program that generated Tor domains with commonly used keywords. This program also checked if the domain is registered in the Tor network. 

Tor relay modification is useful because Tor network structure does not have a public DNS server, it uses a service called Hidden Service DirecTory(HSDir). Some Tor relays are categorized with the flag HSDir. This means that they store information about hidden services. To obtain more valid Tor domains, we deployed a modified Tor relay and got the HSDir flag. This technique was previously reported in ~\cite{noubir2016honey}. We edited the source code of the Tor relay to store hidden service descriptors that identify Tor domains. This technique is key to complete the dataset since it unveils domains that are not published on indexes or linked to other websites.  

\subsection{Dataset coverage}
The process to create the dataset took two years and used the combination of methods previously described. We also combined the results of the i2p and Tor crawling, allowing us to get more domains. We argue that is the reason for which we obtained substantially more domains than DUTA-10k dataset. The dataset stored the cross-relationships between i2p and Tor. 487 relationships in our dataset are references from i2p eepsites to Tor hidden services, and there are 8148 references to eepsites in different Tor hidden services. Even though we gathered i2p and Tor domains for a long time, it is necessary to consider the coverage of the dataset. 

\begin{figure*}[t]
\centering
\includegraphics[scale=0.6]{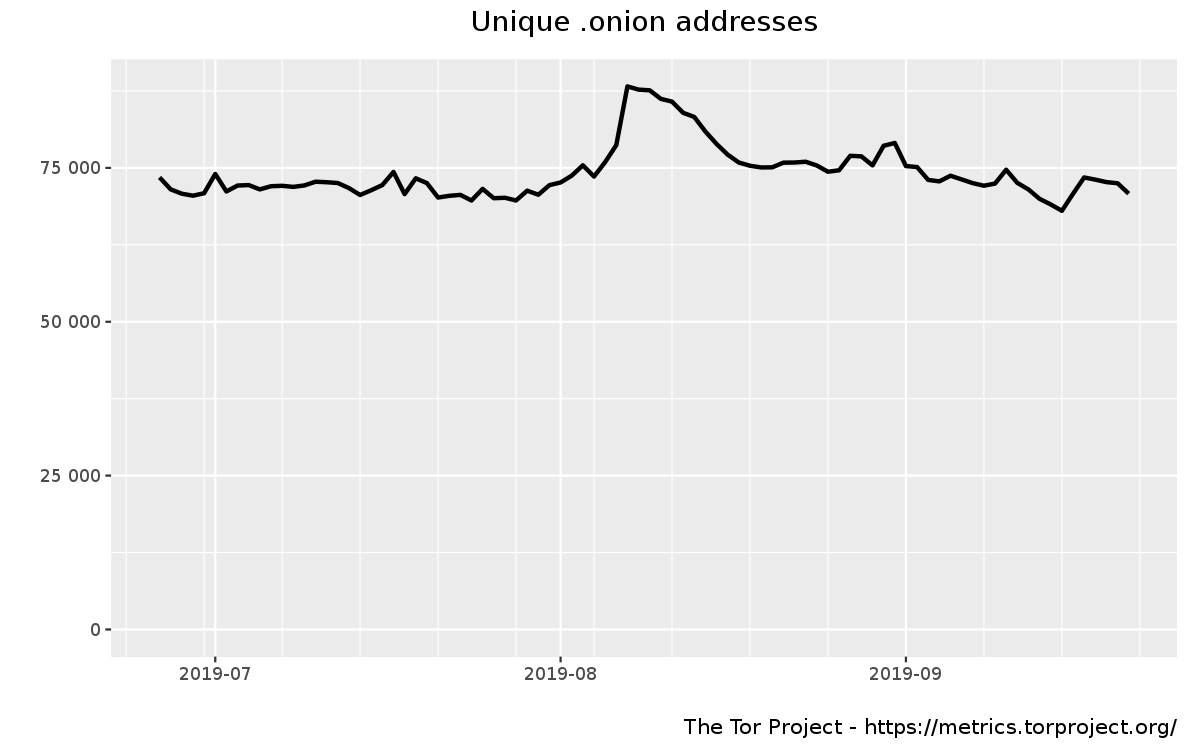}
\caption{Onion services in 2019. Source: metrics.torproject.org (under Creative Commons Attribution 3.0 License - CC BY 3.0 US)}
\label{fig:TorServices}
\end{figure*}

We can estimate the coverage of our dataset by comparing it with the Tor network ~\footnote{https://metrics.torproject.org/}. The Tor project states that there are around 75k unique onion services in 2019 (Figure \ref{fig:TorServices}).

If we compare 75k with the \textgreater 46k onion domains from our dataset, we estimate that it covers around 61\% of the Tor network. However, it is necessary to point out that not all of these hidden services will be websites and that the domains of the dataset may go offline at any moment. Because of this, it is difficult to estimate the exact coverage of Tor darknet, despite the fact that our dataset contains more domains than other investigations. 

Something similar happens if we try to estimate the coverage of the i2p network for our dataset. i2p does not provide official metrics. Although it maintains several updated indexes of domains, many eepsites are published in jump sites and public indexes. Also, as we mentioned before, there is no public i2p dataset of eepsites or research about i2p size or dimension, making difficult to estimate the coverage of i2p of our dataset.

\section{Methodology}
\subsection{Graph construction}
The dataset was used to build a directed graph that represented the network. Nodes represent domains, and edges represent links between domains. All nodes are darknet domains from i2p and Tor. Surface web domains are not included in the dataset and the graph, because our interest is to study the interconnection between darknets. Also, the graph does not include duplicated links (same source and same target), and domains that point to themselves (self-links) as our study focuses on the unweighted connections existing between different nodes.

\subsection{Graph analysis}
Graph analysis is used to compute metrics of graphs, nodes, and their relationships. Many graph algorithms have their origins in Social Network Analysis (SNA), so graphs are sometimes called networks. In this research, we use graph metrics to analyze and compare the darknets (i2p and Tor) and the connections between them. This study reports the following graph metrics: density, average path length, diameter, average degree, and the number of connected components. Density is the ratio of the number of edges present to the total possible number of edges. The average path length is the average distance of all nodes to all other nodes. Diameter is the largest shortest path that can be found between any two nodes in the graph. Average degree measures the average number of incoming and outgoing edges of all nodes. A connected component is a set of connected nodes where each node is reachable from any other node in the same set. The number of connected components is the number of sets that are connected in this way. If all nodes are connected, then the graph contains only one connected component. Similarly, if a node is isolated having no edges to any other node, then it forms a component of just one node. 

Node metrics are used to identify the most important nodes in the darknet. This study reports the following node metrics: degree, closeness centrality, betweenness centrality, and PageRank. Degree is an indicator of the popularity of individual nodes that measures the number of incoming and outgoing links in each node. As both numbers can be different in a directed graph, we distinguish between in-degree and out-degree. Closeness centrality measures the average inverse distance to all other nodes. Higher values of closeness centrality imply shortest distances to all other nodes. Nodes with a high closeness centrality can spread the information very efficiently through the graph. In this study, we use harmonic closeness centrality because it can deal with unconnected graphs. Betweenness centrality is a measure of the amount of influence that a node has over the flow of information in the graph. It is used to find the nodes that serve as bridges between different parts of the graph. Betweenness centrality of a node is computed as the number of shortest paths that pass through the node. Nodes that most frequently lie on these shortest paths have a higher betweenness centrality. PageRank~\cite{brin1998anatomy} is the algorithm used by Google to rank web pages of search results. It is a measure of the transitive influence or connectivity of nodes that considers the number and the quality of the links to the node to determine its relevance. Relevant nodes are likely to receive a higher amount of links from other relevant nodes.

\section{Results}
\subsection{Analysis of darknets}
Table~\ref{tab:metrics} presents the network metrics for i2p, Tor, and for the network that includes the nodes of both. To compute the metrics of this section, we created two subgraphs that included only the nodes in i2p and their connections, and only the nodes in Tor and their connections. We can see that the size of the networks differs significantly. Although Tor has many more nodes and edges, the differences for the average degree, average path length and diameter are not very large. As new domains join the network, they make a proportional number of new links and they remain close to the central nodes. Diameter is small, and density is low in all cases which is common in networks that represent human activities~\cite{chakrabarti2006graph}~\cite{mislove2007measurement}. Metrics of the graph that only contains the Tor nodes and of the graph that contains the nodes of both networks return similar values, except the number of connected components which is reduced significantly when the metric is computed for the complete graph. i2p nodes then increase the connectedness reducing the number of subgraphs (and of nodes) that are unreachable from the main component.

\begin{table}
    \centering
    \caption{Metrics of the networks (graphs) that included only i2p nodes, only Tor nodes, and all nodes}
    \label{tab:metrics}
  \begin{tabular}{c >{\raggedleft\arraybackslash}p{1.5cm} >{\raggedleft\arraybackslash}p{1.2cm} >{\raggedleft\arraybackslash}p{1.2cm}}
    \toprule
    Metric & i2p (eepsites) & Tor (hidden services) & 
    i2p + Tor (eepsites + hidden services)\\
    \midrule
    Nodes & 2687 & 46562 & 49249\\ 
    Edges & 13857 & 282270 & 304673 \\ 
    Avg. Degree & 5.517 & 6.062 & 6.186 \\  
    Density & 0.002 & \textless 0.001 & \textless 0.001\\ 
    Avg. Path Length & 2.769 & 4.356 & 4.412  \\   
    Diameter & 8 & 11 & 12 \\ 
    Connected Components & 11 & 616 & 328\\
  \bottomrule
\end{tabular}
\end{table}

\begin{figure*}[t]
\centering
\includegraphics[scale=1.1]{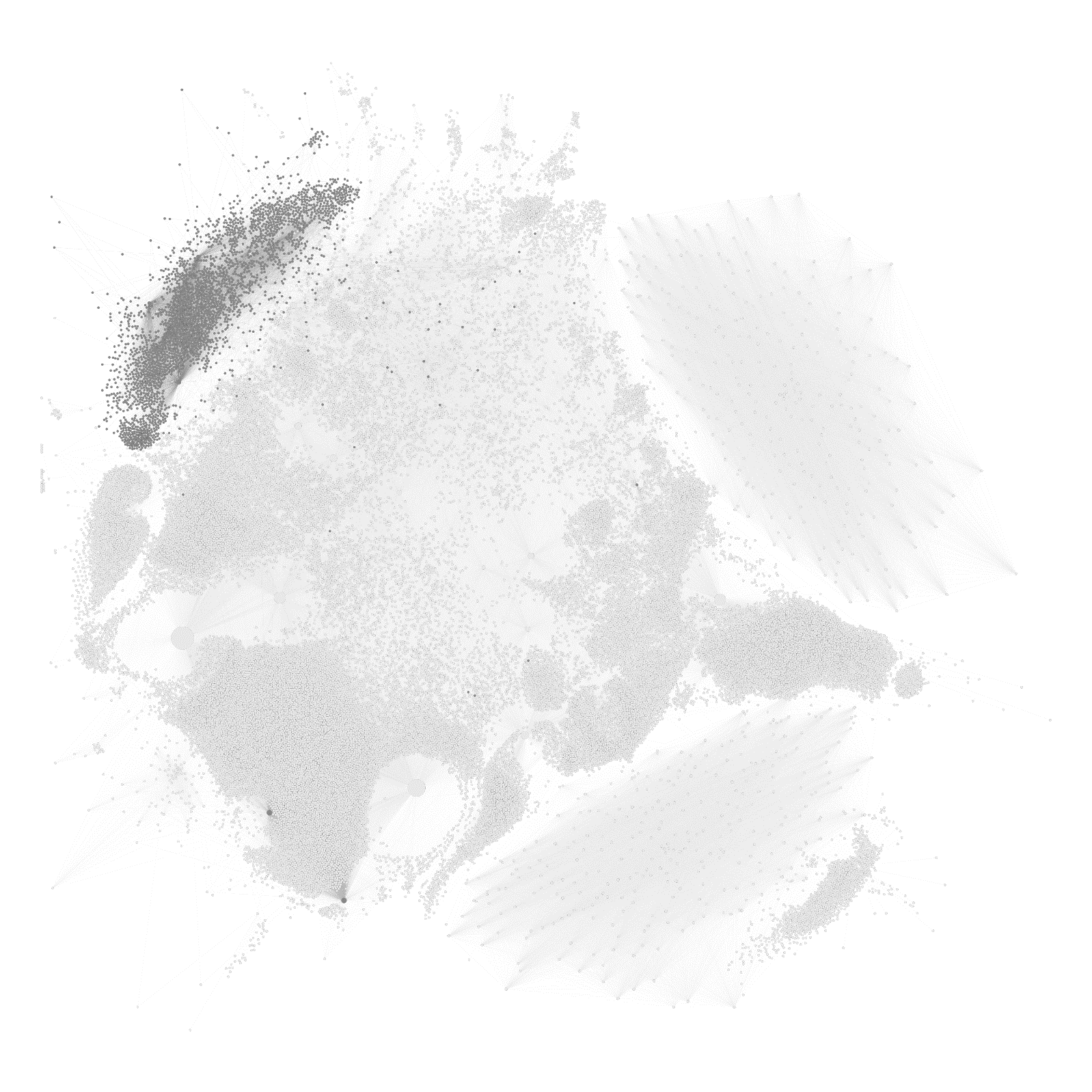}
\caption{Graph of the central component of the Tor+i2p network. Light gray nodes represent Tor domains. Dark gray nodes represent i2p domains. Lines represent links. Size is proportional to the degree (in-degree+out-degree). The figure in high resolution is available at \url{https://gitlab.com/ciberseg-uah/interconection-between-darknets-dataset}}
\label{fig:graph}
\end{figure*}

Figure \ref{fig:graph} presents the graph including all Tor and i2p domains. The size of the nodes is proportional to the number of links (in-degree + out-degree). Only the nodes of the central connected component are included. Although there are many nodes and it is difficult to analyze specific nodes, we can still observe several patterns. Most i2p domains are clustered in the top left side forming a clear subnetwork. Several other i2p nodes are part of the ‘Tor side.’ Particularly, we can see two i2p hosting services (bottom left in the figure) that are primarily linked by Tor domains.

\subsection{Metrics of the domains}
Nodes with the highest in-degree are listed in Table~\ref{tab:degree}. In-degree accounts for the popularity of domains as measured by the number of services that link them. The first domain is incredibly popular, getting close to 50\% of all the possible links. It is the DarkNet Light web that offers multiple Tor links. The out-degree of the second node is also very high since more than 1/3 of all possible services link it. It is a Tor hidden service onion crawler called FreshOnions\footnote{https://github.com/dirtyfilthy/freshonions-torscraper} that crawls the darknet looking for new hidden services and also finds hidden services from Clearnet sources. The remaining top five Tor domains are three Onion Lists of hidden services, like The Onion Crate (ranked 5) which offers a directory with thousands of classified websites from the dark web. The highest i2p node is ranked in position 12. It is followed by andmp.i2p which is Daniel’s Hosting, a free anonymous hosting service that can be found both in i2p and Tor. Since in-degree accounts for the number of services that link a given domain, it makes sense that most popular services are those that link and find other services.

\begin{table}
    \centering
    \caption{Nodes with the highest in-degree}
    \label{tab:degree}
  \begin{tabular}{cccr}
    \toprule
    Rank & Domain & Network & In-Degree\\
    \midrule
    1 & pejjyyh7rhv5ctyu.onion & Tor & 22315\\ 
    2 & zlal32teyptf4tvi.onion & Tor & 16373 \\ 
    3 & onionsnjajzkhm5g.onion & Tor & 10095 \\  
    4 & 44llcbgyt22pwvyq.onion & Tor & 6192\\ 
    5 & cratedvnn5z57xhl.onion & Tor & 5332 \\   
    ... & &  & \\ 
    12 & rv6zugykqdhmwwsuglv7j6...b32.i2p & i2p & 3304 \\   
    13 & andmp.i2p & i2p & 3211 \\ 
  \bottomrule
\end{tabular}
\end{table}

Nodes with the highest out-degree are presented in Table~\ref{tab:outdegree}. Out-degree measures the number of outgoing links of a given domain. The top domain is the i2p proxy service, which is followed by three different jump services from i2p (ranked 2-4). Since i2p jump services act as second layer proxies between client and host when the client does not have the address in its i2p ‘addressbook’, it makes sense that these services are then the major hubs of the hidden networks providing links to more services than any other. Tor does not offer similar services and the findability of onions is limited mostly to lists. So results suggest that i2p proxy and jump services are an important entry point to the network. Daniel’s hosting service from Tor is ranked fifth and it is the service with the highest out-degree of this darknet. The Tor mailbox service follows (ranked sixth). Still, the values of the out-degree for top-ranked nodes is much less when compared with in-degree. We can also observe that in-degree does not match out-degree in both rankings and also that the top-ranked services are different. In-degree and out-degree differ considerably for each node meaning that domains are either a provider of links or the subject of them. The low out-degree of the Tor services as compared to in-degree shows that this darknet is highly distributed when it comes to finding onions. The hubs of Tor are a hosting service and a mail service.

\begin{table}
    \centering
    \caption{Nodes with the highest out-degree}
    \label{tab:outdegree}
  \begin{tabular}{cccr}
    \toprule
    Rank & Domain & Network & Out-Degree\\
    \midrule
    1 & proxy.i2p & i2p & 1793\\ 
    2 & stats.i2p & i2p & 1205 \\ 
    3 & no.i2p & i2p & 1185 \\  
    4 & i2pjump.i2p & i2p & 1177\\ 
    5 & dhosting4xxoydyaiv...syd.onion & Tor & 535 \\   
    6 & Torbox3uiot6wchz.onion & Tor & 337 \\
  \bottomrule
\end{tabular}
\end{table}


Top-ranked nodes in terms of closeness centrality (Table~\ref{tab:closeness}) are Tor domains. The highest-ranked i2p node is in position 27, while the second i2p node is in position 967. Nodes with the highest in-degree also occupy top positions in terms of closeness centrality, suggesting that services massively pointed are in the central positions of the network being able to spread information efficiently.  The top four domains are the same domains of the in-degree ranking. These include three Tor onion lists and the Tor hidden service onion crawler (FreshOnions). The Undernet Directory (UnderDir) completes the top 5. UnderDir is another onion list that classifies links by language and topic. The top-ranked i2p domain is Daniel’s Hosting service. Closeness centrality is normalized, and the maximum possible value is 1. A value of 1 means that the node is connected to all other nodes by the minimum possible distance. Top-ranked nodes have very high values suggesting that they are very closely connected to all other nodes. So most of the nodes are directly linked to Onion lists or the FreshOnion crawler. Since these sites contain lists of services, results suggest that this information has a prominent and central position in the hidden networks.

\begin{table}
    \centering
    \caption{Nodes with the highest closeness harmonic centrality}
    \label{tab:closeness}
  \begin{tabular}{cccr}
    \toprule
    Rank & Domain & Network & Closeness\\
    \midrule
    1 & pejjyyh7rhv5ctyu.onion & Tor & 0.706\\ 
    2 & zlal32teyptf4tvi.onion & Tor & 0.642 \\ 
    3 & onionsnjajzkhm5g.onion & Tor & 0.580 \\  
    4 & 44llcbgyt22pwvyq.onion & Tor & 0.559\\ 
    5 & underdj5ziov3ic7.onion & Tor & 0.513\\ 
    ... & &  & \\ 
    27 & andmp.i2p & i2p & 0.470 \\
  \bottomrule
\end{tabular}
\end{table}

Results of betweenness centrality are presented in Table~\ref{tab:betweenness}. Top-ranked nodes include two domains from i2p and four domains from Tor. The node with the highest betweenness centrality is an i2p jump service ’i2pjump.i2p’. Rank second is the FreshOnions Tor service which is the only top-ranked crawler. Daniel’s Hosting Tor service ranks third in terms of betweenness centrality. Rank fourth is the DarkNet Light Tor website, and rank fifth is another onion list. In the sixth position, we find the second i2p domain, ‘Hiddenanswers.i2p.’ Hiddenanswers is a question and answer website that is also present in Tor. The fact that domains from both networks are highly ranked reflects the importance of i2p domains as brokerage agents in the networks. The reduction of the number of connected components when both networks are analyzed together, previously mentioned, also supports this argument. The variety of services (jump service and Q\&A from i2p, as well as crawler, hosting and onion lists from Tor) also suggest that all play a significant role in the darknet ecosystem. Betweenness centrality is a measure of the number of shortest paths that go through a given node, and paths represent sequences of links that users follow in the darknets. So these particular nodes are in brokerage positions for the kind of service that they offer (jump, crawler, hosting, and onion list) in the paths that users follow and in the flows of information in the darknet.

\begin{table}
    \centering
    \caption{Nodes with the highest betweenness centrality}
    \label{tab:betweenness}
  \begin{tabular}{cccr}
    \toprule
    Rank & Domain & Network & Betweenness\\
    \midrule
    1 &	i2pjump.i2p &	i2p	& 43755077\\ 
    2 &	zlal32teyptf4tvi.onion &	Tor &	41176352\\  
    3 &	dhosting4xxoydyaiv...syd.onion &	Tor &	32290935\\ 
    4 &	pejjyyh7rhv5ctyu.onion &	Tor &	28768547\\ 
    5 &	onionsnjajzkhm5g.onion &	Tor &	27588772\\ 
    6 &	hiddenanswers.i2p &	i2p &	24853420\\
  \bottomrule
\end{tabular}
\end{table}

Table~\ref{tab:pagerank} presents the nodes with the highest PageRank. Tor domains occupy top positions. The highest i2p domain is in position ten. PageRank measures the importance of a node in terms of the number and quality of the links to it. It returns the probability that a user randomly clicking on links will arrive to a site. Results show that Tor domains are more important than i2p domains. We can see that results are similar to other metrics as domains ranked in positions two (DarkNet Light), three (FreshOnions), and four (an onion list) are also highly ranked in terms of in-degree, closeness and betweenness. This suggests that the number of incoming links is also a good indicator of the position and influence of a domain. Ranked in the fifth position is another Tor onion list. However, when it comes to PageRank, the top domain is Daniel’s Hosting, which also has a high in-degree (rank 17), out-degree (rank 5) and betweenness (rank 3). Since the PageRank score is substantially higher for the Tor hosting site, results suggest that this site is the most important domain and it will likely receive more visits than any other. DarkNet Light is the most important list of onion services in Tor, and it will likely receive many more visits than other onion lists that follow it in the ranking. identiguy.i2p is the top-ranked i2p domain (rank 10). It is an i2p jump service similar to the stats.i2p jump site.

\begin{table}
    \centering
    \caption{Nodes with the highest PageRank centrality}
    \label{tab:pagerank}
  \begin{tabular}{cccr}
    \toprule
    Rank & Domain & Network & PageRank\\
    \midrule
    1 &	dhosting4xxoydyaiv...syd.onion &	Tor	& 3492\\
    2 &	pejjyyh7rhv5ctyu.onion &	Tor &	2897\\  
    3 &	zlal32teyptf4tvi…syd.onion &	Tor &	1478\\ 
    4 &	onionsnjajzkhm5g.onion &	Tor &	1129\\ 
    5 &	donionsixbjtiohce2...ead.onion &	Tor &	1077\\ 
    ... &	& &	\\ 
    10 & identiguy.i2p	& i2p & 778	\\ 
  \bottomrule
\end{tabular}
\end{table}

\section{Discussion}
Results show that the most popular Tor websites are focused on indexing domains. Tor network does not have a public index with all the hidden domains, so several websites just maintain lists of popular and new domains. Hence, to further index and study the hidden services in Tor, we should focus first on the top-ranked domains for the different metrics. These are the most critical sites to start mapping the darknets. 

As for i2p, the most important domains are jump and proxy sites. Due to the way i2p works, normally, these sites are the most relevant. To access eepsites, it is necessary to use a jump site. Jump sites act as a DNS service to access i2p domains. Hence, if we want to index and study i2p network, the eepsites that are top-ranked in out-degree and betweenness centrality should guide the investigation. 

Results also show that other types of sites have prominent positions in each darknet. In Tor, one of the services with the highest out-degree is TorBox mail service, suggesting that this Tor service is used mostly to point to other hidden services. This may be an indicator of the fact that a lot of illegal activities use TorBox as an email server~\cite{yetter2015darknets}~\cite{dolliver2019emerging}~\cite{gross2019moniToring}. As for i2p, results for betweenness centrality suggest that \textit{hiddenanswers.i2p} plays a significant role in the communication paths of the network. Hiddenanswers is a site with similar functionality to Yahoo Answers\footnote{\url{https://answers.yahoo.com/}} but without any kind of censorship. Hiddenanswers is also present in Tor. The duplicity of this as well as other services further strengthens the assumption of the interconnection and the dual nature of the communication between Tor and i2p.

Previous research focuses on describing or improving the security of Tor~~\cite{mccoy2008shining}~~\cite{snader2008tune}. Several datasets of Tor have been reported and analyzed, like DUTA-10K~\cite{ALNABKI2019212}. We compared our dataset with it, showing that our approach provides a broader coverage of Tor. Further, our work also recorded i2p eepsites, reporting the connections between Tor and i2p. To the best of our knowledge, there are no datasets or studies that systematically map i2p structure. Research on i2p focuses on monitoring and attacking i2p users~\cite{egger2013practical}~\cite{timpanaro2011moniToring}. Our results provide a map of both darknets and their connections which is used to find the most relevant sites in terms of different network metrics. The dataset is publicly accessible in GitLab at  \url{https://gitlab.com/ciberseg-uah/interconection-between-darknets-dataset}.

\section{Scenarios of use for LEAs}
The dataset and findings presented in this paper can help LEAs to investigate darknet illegal activities in different ways. First of all, this paper evidences the existence of cross operations between i2p and Tor, stressing the necessity to investigate i2p discover Tor domains. Past operations of LEAs mostly focused on Tor~\cite{noauthor_double_nodate}~\cite{lacson201621st}, and because of this, cybercriminals are probably moving to i2p. LEAs need to be aware of this and focus on other darknets, since they offer different and complementary services. LEAs can also use the dataset to investigate current Tor and i2p domains. They can find existing domains and track their connections. They can also run new analyses to investigate the positioning and influence of services. Furthermore, our analysis of the dataset provides a list of current key sites of Tor and i2p. LEAs can use it as a starting point to crawl and monitor the darknets using the methods presented in section III.A. LEAs could then create new mappings that include services potentially leading to new criminal activity. Knowing the main types of services in the darknet ecosystem could guide future investigations. This is particularly important because darknet services are highly volatile. Even if a given domain goes offline, it is likely that it will be replaced by other offering similar functionalities. As hidden services can appear and disappear relatively quickly, crawling the darknet at a given moment provides a snapshot that LEAs can use to monitor suspicious activity. These can be improved in the future to create dynamic views of the Darknet or of a part of it, that may even raise alerts when given events happen. For instance, when a service comes online.

Another possible scenario in which the dataset can be useful is when LEAs try to take down criminal sites. This often involves performing some kind of cyberattack over them, which usually is the only option due to the built-in opacity of darknets. Discovering as much as possible of the attack surface of an objective is crucial to find weaknesses that can be used to define attack vectors. The dataset can help to find services that are present in Tor and i2p. There is the possibility that a service can be securely deployed in Tor but not in i2p or the other way around. LEAs can try to attack the i2p exposed service to take down or take control of the site~\cite{van2018lost}~\footnote{Alphabay and Hansa darknet markets shut down after international police operation. \url{https://www.dw.com/en/alphabay-and-hansa-darknet-markets-shut-down-after-international-police-operation/a-39776885}}. Our analysis of the dataset already pointed to prominent sites that are mirrored in Tor and i2p (like hosting services) showing how they can be found. It also suggests the necessity of finding all the access points of criminal services in the Darknet to expose their complete attack surface and bring them down for good. LEAs can also analyze the current dataset to find other services that may be mirrored.

\section{Conclusion}
This paper presents the results of analyzing a dataset with data of i2p and Tor networks. We also report the process of creating the dataset. To the best of our knowledge, this is the only dataset that connects eepsites and hidden services. 

Using graph analysis, we showed that crawling one darknet can improve the discovery of sites present on a different darknet. Also, this study helps to understand the positioning of sites and their influence in the network. 

Graph metrics return similar values for most measures suggesting that both darknets are structurally similar. The reduction of the number of connected components, when all nodes are included in a single graph, suggests that i2p nodes play an important role in making more Tor nodes reachable.

Node metrics suggest that in terms of centrality, Tor onion lists and the Tor FreshOnions crawler occupy central positions and get most of the links playing a substantial role in the spreading information. i2p jump services are the main hubs of the network (top-ranked out-degree) pointing to many other nodes. Results of betweenness centrality suggest that all main darkenet services (jump, crawl, hosting, and onion lists) play an important role in the communication paths of the darknets. The most significant actors for each role are i2pjump.i2p, the FreshOnions Tor crawler, Daniel's Tor hosting, and the DarkNet Light Tor website (onion list).

The most important Tor domains are hosting and index websites. Furthermore, the most important i2p domains are jump sites. Since Tor does not offer effective search engines, index websites partially address this limitation. So i2p jump sites become the relevant hubs of the darknet.


\bibliographystyle{IEEEtran}
\bibliography{main_ieee}

\begin{thebibliography}{10}
\providecommand{\url}[1]{#1}
\csname url@samestyle\endcsname
\providecommand{\newblock}{\relax}
\providecommand{\bibinfo}[2]{#2}
\providecommand{\BIBentrySTDinterwordspacing}{\spaceskip=0pt\relax}
\providecommand{\BIBentryALTinterwordstretchfactor}{4}
\providecommand{\BIBentryALTinterwordspacing}{\spaceskip=\fontdimen2\font plus
\BIBentryALTinterwordstretchfactor\fontdimen3\font minus
  \fontdimen4\font\relax}
\providecommand{\BIBforeignlanguage}[2]{{%
\expandafter\ifx\csname l@#1\endcsname\relax
\typeout{** WARNING: IEEEtran.bst: No hyphenation pattern has been}%
\typeout{** loaded for the language `#1'. Using the pattern for}%
\typeout{** the default language instead.}%
\else
\language=\csname l@#1\endcsname
\fi
#2}}
\providecommand{\BIBdecl}{\relax}
\BIBdecl

\bibitem{chertoff_impact_2015}
\BIBentryALTinterwordspacing
M.~Chertoff and T.~Simon, ``\BIBforeignlanguage{en}{The {Impact} of the {Dark}
  {Web} on {Internet} {Governance} and {Cyber} {Security}},'' Feb. 2015.
  [Online]. Available:
  \url{https://www.cigionline.org/publications/impact-dark-web-internet-governance-and-cyber-security}
\BIBentrySTDinterwordspacing

\bibitem{owen_Tor_2015}
\BIBentryALTinterwordspacing
G.~Owen and N.~Savage, ``The {Tor} {Dark} {Net},'' Centre for International
  Governance Innovation and the Royal Institute of International Affairs, Sep.
  2015. [Online]. Available:
  \url{https://www.cigionline.org/publications/tor-dark-net}
\BIBentrySTDinterwordspacing

\bibitem{snader2008tune}
R.~Snader and N.~Borisov, ``A tune-up for tor: Improving security and
  performance in the tor network.'' in \emph{ndss}, vol.~8, 2008, p. 127.

\bibitem{mccoy2008shining}
D.~McCoy, K.~Bauer, D.~Grunwald, T.~Kohno, and D.~Sicker, ``Shining light in
  dark places: Understanding the tor network,'' in \emph{International
  symposium on privacy enhancing technologies symposium}.\hskip 1em plus 0.5em
  minus 0.4em\relax Springer, 2008, pp. 63--76.

\bibitem{loesing2010case}
K.~Loesing, S.~J. Murdoch, and R.~Dingledine, ``A case study on measuring
  statistical data in the tor anonymity network,'' in \emph{International
  Conference on Financial Cryptography and Data Security}.\hskip 1em plus 0.5em
  minus 0.4em\relax Springer, 2010, pp. 203--215.

\bibitem{dolliver2015evaluating}
D.~S. Dolliver, ``Evaluating drug trafficking on the tor network: Silk road 2,
  the sequel,'' \emph{International Journal of Drug Policy}, vol.~26, no.~11,
  pp. 1113--1123, 2015.

\bibitem{wilson2016forensic}
M.~Wilson and B.~Bazli, ``Forensic analysis of i2p activities,'' in \emph{2016
  22nd International Conference on Automation and Computing (ICAC)}.\hskip 1em
  plus 0.5em minus 0.4em\relax IEEE, 2016, pp. 529--534.

\bibitem{bazli2017dark}
B.~Bazli, M.~Wilson, and W.~Hurst, ``The dark side of i2p, a forensic analysis
  case study,'' \emph{Systems Science \& Control Engineering}, vol.~5, no.~1,
  pp. 278--286, 2017.

\bibitem{nunes2016darknet}
E.~Nunes, A.~Diab, A.~Gunn, E.~Marin, V.~Mishra, V.~Paliath, J.~Robertson,
  J.~Shakarian, A.~Thart, and P.~Shakarian, ``Darknet and deepnet mining for
  proactive cybersecurity threat intelligence,'' in \emph{2016 IEEE Conference
  on Intelligence and Security Informatics (ISI)}.\hskip 1em plus 0.5em minus
  0.4em\relax IEEE, 2016, pp. 7--12.

\bibitem{chaabane2010digging}
A.~Chaabane, P.~Manils, and M.~A. Kaafar, ``Digging into anonymous traffic: A
  deep analysis of the tor anonymizing network,'' in \emph{2010 Fourth
  International Conference on Network and System Security}.\hskip 1em plus
  0.5em minus 0.4em\relax IEEE, 2010, pp. 167--174.

\bibitem{ALNABKI2019212}
\BIBentryALTinterwordspacing
M.~W. Al-Nabki, E.~Fidalgo, E.~Alegre, and L.~Fernández-Robles, ``Torank:
  Identifying the most influential suspicious domains in the tor network,''
  \emph{Expert Systems with Applications}, vol. 123, p. 212 – 226, 2019.
  [Online]. Available:
  \url{http://www.sciencedirect.com/science/article/pii/S0957417419300296}
\BIBentrySTDinterwordspacing

\bibitem{iliou2016hybrid}
C.~Iliou, G.~Kalpakis, T.~Tsikrika, S.~Vrochidis, and I.~Kompatsiaris, ``Hybrid
  focused crawling for homemade explosives discovery on surface and dark web,''
  in \emph{2016 11th International Conference on Availability, Reliability and
  Security (ARES)}.\hskip 1em plus 0.5em minus 0.4em\relax IEEE, 2016, pp.
  229--234.

\bibitem{goodreau2007advances}
S.~M. Goodreau, ``Advances in exponential random graph (p*) models applied to a
  large social network,'' \emph{Social networks}, vol.~29, no.~2, pp. 231--248,
  2007.

\bibitem{sanchez2017onions}
I.~Sanchez-Rola, D.~Balzarotti, and I.~Santos, ``The onions have eyes: A
  comprehensive structure and privacy analysis of tor hidden services,'' in
  \emph{Proceedings of the 26th International Conference on World Wide
  Web}.\hskip 1em plus 0.5em minus 0.4em\relax International World Wide Web
  Conferences Steering Committee, 2017, pp. 1251--1260.

\bibitem{noubir2016honey}
G.~Noubir and A.~Sanatinia, ``Honey onions: Exposing snooping tor hsdir
  relays,'' in \emph{DEF CON 24}, 2016.

\bibitem{brin1998anatomy}
S.~Brin and L.~Page, ``The anatomy of a large-scale hypertextual web search
  engine,'' \emph{Computer networks and ISDN systems}, vol.~30, no. 1-7, pp.
  107--117, 1998.

\bibitem{chakrabarti2006graph}
D.~Chakrabarti and C.~Faloutsos, ``Graph mining: Laws, generators, and
  algorithms,'' \emph{ACM computing surveys (CSUR)}, vol.~38, no.~1, p.~2,
  2006.

\bibitem{mislove2007measurement}
A.~Mislove, M.~Marcon, K.~P. Gummadi, P.~Druschel, and B.~Bhattacharjee,
  ``Measurement and analysis of online social networks,'' in \emph{Proceedings
  of the 7th ACM SIGCOMM conference on Internet measurement}.\hskip 1em plus
  0.5em minus 0.4em\relax ACM, 2007, pp. 29--42.

\bibitem{yetter2015darknets}
R.~B. Yetter, ``Darknets, cybercrime \& the onion router: Anonymity \&security
  in cyberspace,'' Ph.D. dissertation, Utica College, 2015.

\bibitem{dolliver2019emerging}
D.~S. Dolliver, ``Emerging technologies, law enforcement responses, and
  national security,'' \emph{ISJLP}, vol.~15, p. 123, 2019.

\bibitem{gross2019moniToring}
W.~F. Gross~Jr, ``Monitoring and tracking isis on the dark web,'' \emph{Online
  Terrorist Propaganda, Recruitment, and Radicalization}, p. 341, 2019.

\bibitem{egger2013practical}
C.~Egger, J.~Schlumberger, C.~Kruegel, and G.~Vigna, ``Practical attacks
  against the i2p network,'' in \emph{International Workshop on Recent Advances
  in Intrusion Detection}.\hskip 1em plus 0.5em minus 0.4em\relax Springer,
  2013, pp. 432--451.

\bibitem{timpanaro2011moniToring}
\BIBentryALTinterwordspacing
J.~P. Timpanaro, C.~Isabelle, and F.~Olivier, ``Monitoring the i2p network,''
  2011. [Online]. Available: \url{https://hal.inria.fr/hal-00653136/document}
\BIBentrySTDinterwordspacing

\bibitem{noauthor_double_nodate}
\BIBentryALTinterwordspacing
``\BIBforeignlanguage{en}{Double blow to dark web marketplaces}.'' [Online].
  Available:
  \url{https://www.europol.europa.eu/newsroom/news/double-blow-to-dark-web-marketplaces}
\BIBentrySTDinterwordspacing

\bibitem{lacson201621st}
W.~Lacson and B.~Jones, ``The 21st century darknet market: Lessons from the
  fall of silk road.'' \emph{International Journal of Cyber Criminology},
  vol.~10, no.~1, 2016.

\bibitem{van2018lost}
R.~van Wegberg and T.~Verburgh, ``Lost in the dream? measuring the effects of
  operation bayonet on vendors migrating to dream market,'' in
  \emph{Proceedings of the Evolution of the Darknet Workshop}, 2018, pp. 1--5.

\end{thebibliography}

\section*{Authors' Biographies}

\subsection*{}
\begin{wrapfigure}{l}{25mm} 
    \includegraphics[width=1in,height=1.25in,clip,keepaspectratio]{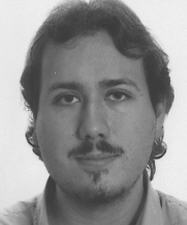}
\end{wrapfigure}\par
\textbf{Carlos Cilleruelo} holds B.Sc (2015) in Computer Science and M.Sc. in Cyber Security (2016). Currently he works as a researcher at the University of Alcalá (2018-) where he is also completing his Ph.D. (2018-). Previously, he worked in several companies as a cybersecurity and forensics specialist. He also participated in an Erasmus+ program on data forensics (2013). He is a research member of the ProTego EU H2020 project focused in cybersecurity and e-health (2019-2021). He presented the results of his work at the RootedCon-2020 conference. His research interest focus on cybersecurity, particularly in the areas of forensics, darknets and machine learning applied to cybersecurity.\par

\subsection*{}
\begin{wrapfigure}{l}{25mm} 
    \includegraphics[width=1in,height=1.25in,clip,keepaspectratio]{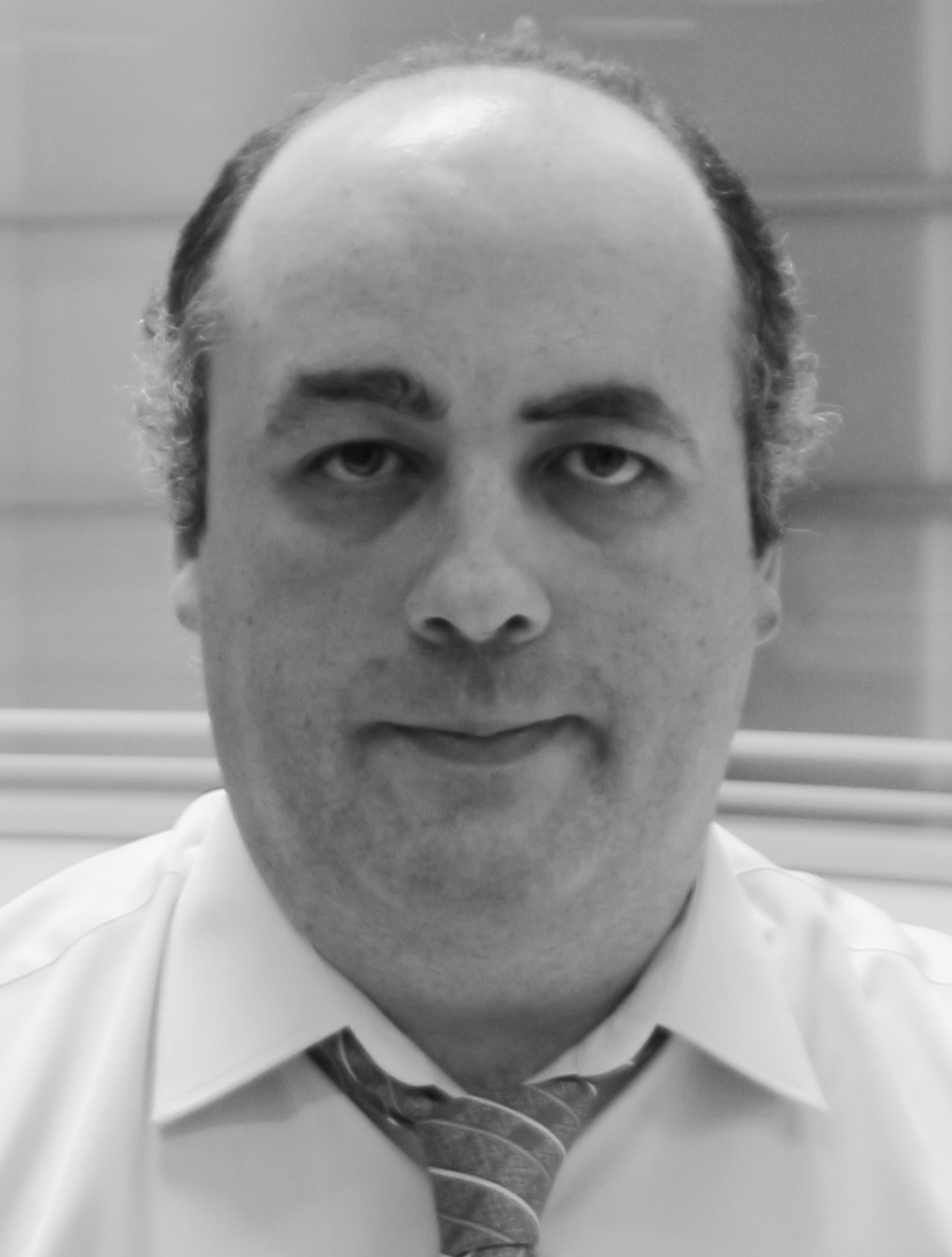}
\end{wrapfigure}\par
\textbf{PhD. Luis de-Marcos} received the B.Sc. and M.Sc. degrees in computer science and the Ph.D. degree in information, documentation, and knowledge in 2001, 2005, and 2009, respectively. He has been an Associate Professor with the University of Alcalá since 2015. He is Principal Investigator of the Research Team of the ProTego Project (H2020) on Cybersecurity (2019–2021), and was a Principal Investigator in two national research projects in Spain (2011–2013). He was a Research Fellow with Lund University, Sweden, in 2007 and 2009, the University of Reading, U.K., in 2008, the Monterrey Institute of Technology, Mexico, in 2010, and the University of Zagreb, Croatia, in 2018. He has over 100 refereed publications in conferences and journals. His research interests include educational technologies, e-learning and cybersecurity.\par

\subsection*{}
\begin{wrapfigure}{l}{25mm} 
    \includegraphics[width=1in,height=1.25in,clip,keepaspectratio]{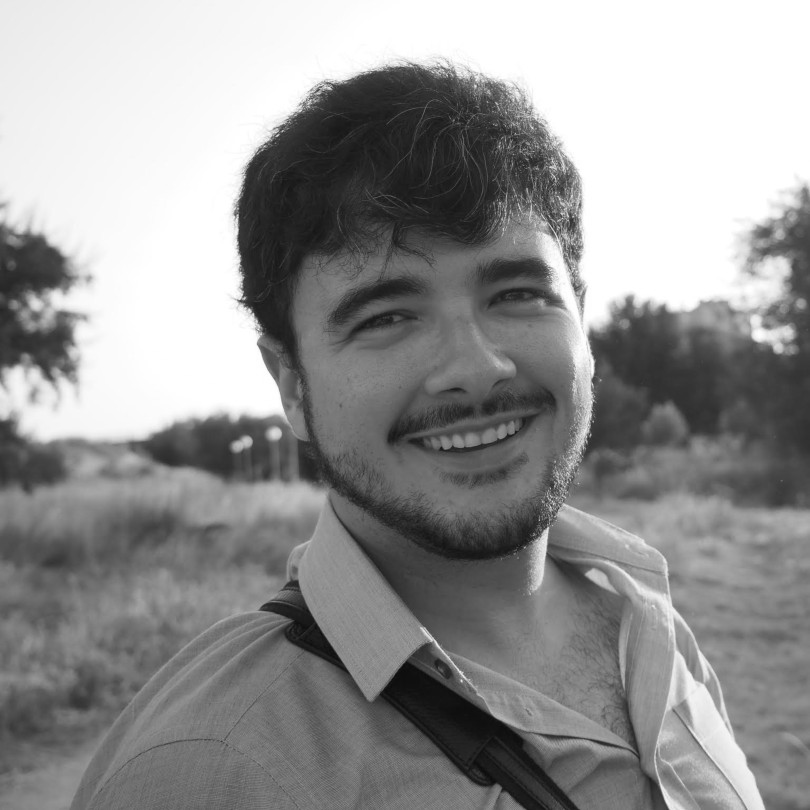}
\end{wrapfigure}\par
\textbf{Javier Junquera-Sánchez}  holds B.Sc. in Computer Science (2018) and M.Sc. in Cyber Security (2019). Currently he works as researcher at the University of Alcalá (2018-) where he is also  a PhD student (2020-). Previously, he worked for several companies as cybersecurity analysts and software developer.He is a research member of the ProTego EU H2020 project focused in cybersecurity and e-health (2019-2021). He presented the results of his work at the RootedCon-2020 conference. His research interests focus on cybersecurity, software security, cryptography and darknets.\par

\subsection*{}
\begin{wrapfigure}{l}{25mm} 
    \includegraphics[width=1in,height=1.25in,clip,keepaspectratio]{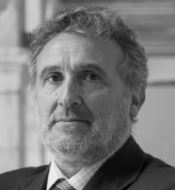}
\end{wrapfigure}\par
\textbf{PhD. José Javier Martínez-Herráiz}   obtained his degree in Computer Science from the Polytechnic University of Madrid and his PhD from the University of Alcalá (2004). He has been Associate Professor at the Department of Computer Science (Artificial Intelligence Area) of the University of Alcalá (1994-), where he is the Rector's Delegate for Electronic Administration and Security. He worked in private telecommunication business companies (Spain and Italy) as software analyst, project manager and consultant between 1988 and 1999.  He was director of the Computer Science Department between 2008 and 2011. He has collaborated extensively with Spanish law enforcement agencies and cybersecurity companies, and his work was awarded with Order of Merit of the Police Forces (2011 and 2015). He has practical experience in software development, technology and modelling, methodologies for software projects, planning and management, software maintenance, e-learning, gamification technology and cybersecurity. \par

\end{document}